\documentclass[prl,twocolumn,preprintnumbers,superscriptaddress,nofootinbib]{revtex4} 

\usepackage{graphicx}
\usepackage{bm}
\usepackage{amsmath}
\usepackage{amssymb}
\usepackage{amsfonts}
\usepackage{fancyhdr}
\usepackage{slashed}
\usepackage{latexsym,epsfig,bbm}

\newcommand{\bra}[1]{\langle{#1}|}
\newcommand{\ket}[1]{|{#1}\rangle}

\newcommand{\braket}[2]{\langle{#1}|{#2}\rangle}
\newcommand{\bopk}[3]{\langle{#1}|{#2}|{#3}\rangle}

\newcommand{\figref}[1]{Fig.~\ref{#1}}

\newcommand{\Tr}{\mathrm{Tr}}
\usepackage{color}
\definecolor{blue}{rgb}{0,0.2,1}

\definecolor{red}{rgb}{0.9,0,0}

\newcommand{\past}[1]{\overleftarrow{#1}}
\newcommand{\fut}[1]{\overrightarrow{#1}}
\newcommand{\pastfut}[1]{\overleftrightarrow{#1}}

\newcommand{\npast}{\overleftarrow{n}}

\newcommand{\nfut}{\overrightarrow{n}}
\newcommand{\Nfut}{\overrightarrow{N}}

\newcommand\blfootnote[1]{
	\begingroup
	\renewcommand\thefootnote{}\footnote{#1}
	\addtocounter{footnote}{-1}
	\endgroup}

\begin{document}

\title{Extreme dimensionality reduction with quantum modelling}

\author{Thomas J.~Elliott$^{\S}$}
\email{physics@tjelliott.net}
\affiliation{Department of Mathematics, Imperial College London, London SW7 2AZ, United Kingdom}
\affiliation{Complexity Institute, Nanyang Technological University, Singapore 637335}
\affiliation{Nanyang Quantum Hub, School of Physical and Mathematical Sciences, Nanyang Technological University, Singapore 637371}
\author{Chengran Yang$^{\S}$}
\email{yangchengran92@gmail.com}
\affiliation{Nanyang Quantum Hub, School of Physical and Mathematical Sciences, Nanyang Technological University, Singapore 637371}
\affiliation{Complexity Institute, Nanyang Technological University, Singapore 637335}
\author{Felix C.~Binder}
\affiliation{Institute for Quantum Optics and Quantum Information, Austrian Academy of Sciences, Boltzmanngasse 3, Vienna 1090, Austria}
\author{Andrew J.~P.~Garner}
\affiliation{Institute for Quantum Optics and Quantum Information, Austrian Academy of Sciences, Boltzmanngasse 3, Vienna 1090, Austria}
\affiliation{Centre for Quantum Technologies, National University of Singapore, 3 Science Drive 2, Singapore 117543}
\author{Jayne Thompson}
\affiliation{Centre for Quantum Technologies, National University of Singapore, 3 Science Drive 2, Singapore 117543}
\author{Mile Gu}
\email{mgu@quantumcomplexity.org}
\affiliation{Nanyang Quantum Hub, School of Physical and Mathematical Sciences, Nanyang Technological University, Singapore 637371}
\affiliation{Complexity Institute, Nanyang Technological University, Singapore 637335}
\affiliation{Centre for Quantum Technologies, National University of Singapore, 3 Science Drive 2, Singapore 117543}

\date{\today}

\begin{abstract}

Effective and efficient forecasting relies on identification of the relevant information contained in past observations -- the predictive features -- and isolating it from the rest. When the future of a process bears a strong dependence on its behaviour far into the past, there are many such features to store, necessitating complex models with extensive memories. Here, we highlight a family of stochastic processes whose minimal classical models must devote unboundedly many bits to tracking the past. For this family, we identify quantum models of equal accuracy that can store all relevant information within a single two-dimensional quantum system (qubit). This represents the ultimate limit of quantum compression and highlights an immense practical advantage of quantum technologies for the forecasting and simulation of complex systems.

\end{abstract}
\maketitle

Predicting the future based on past events is a cornerstone of life. From meteorologists forecasting the weather, through investors trading on stock markets, to a predator chasing its prey, the ability to identify causes and accurately anticipate effects is central to survival and success. To carry out these essential tasks, models must be formulated, and information about past observations must be stored within memory.\blfootnote{\hspace{-1em}$^{\S}$These authors contributed equally to the results.}

In this context, processes with long historical dependence typically require models that store extensive information about past observations. This is because a model must ascribe each set of past causes that can give rise to distinct future effects to distinct configurations in its memory. When there are many such causes, the memory must support many configurations. Classically, the number of configurations is synonymous with the \mbox{\emph{dimension}} of the memory -- tracking a process with causes reaching far into the past typically requires a large memory with many dimensions.

In contrast, the number of configurations a quantum memory can take is separate from its dimension. This has lead to quantum encodings with reduced memory dimension for several Markovian processes -- where each output is conditional only on its immediate predecessor~\cite{thompson2018causal, loomis2019strong, liu2019optimal, ghafari2019dimensional}. Here, we demonstrate that not only do these quantum advantages persist for non-Markovian processes, but that they become even more pronounced in this regime. We consider a family of such processes where the memory dimension required of a faithful classical model diverges with precision, and identify corresponding quantum models that compress all configurations into two dimensions. This allows for all relevant history to be stored in a single two-state quantum system (\emph{qubit}), evincing an extreme quantum advantage that scales without bound. Moreover, our protocol requires only a single probe qubit to extract the future statistics. This turns a problem from the converse scenario -- that tracking a finite quantum system can require infinite classical resources~\cite{monras2016quantum, cabello2016thermodynamical, cabello2018optimal, warszawski2019open} -- into a useful tool.

This complements recent advances at the interface of complexity and quantum science, where it has been found that quantum models can drastically reduce the amount of past information -- as measured by \mbox{\emph{information entropy}} -- that must be stored in memory to replicate the future behaviour of a process~\cite{gu2012quantum, mahoney2016occam, palsson2017experimentally, garner2017provably, aghamohammadi2017extreme, elliott2018superior, thompson2018causal, elliott2019memory}. Our work indicates that this advantage (along with its quantitative scaling divergences) also persists for the \mbox{\emph{memory dimension}}. Crucially, this brings practical, verifiable, and significant quantum memory advantages within the reach of present technologies. 

{\bf Framework and tools.}
A stochastic process $\mathcal{X}$ can be characterised by an observation sequence $\pastfut{X}$, detailing what happens and when~\cite{khintchine1934korrelationstheorie}. We can partition this sequence in two: a past $\past{x}$ that describes everything that has happened up to the present; and a future $\fut{x}$ describing everything yet to come (we use upper case to denote random variables, and lower case for their corresponding variates). The goal of causal modelling is to use the past (and only the past) to simulate the future~\cite{crutchfield1989inferring, shalizi2001computational, crutchfield2012between, thompson2018causal}. Specifically, a \emph{causal model} $\mathsf{M}$ stores in its memory states $m\in \mathcal{M}$ determined from an encoding function of the past \mbox{$f:\{\past{x}\}\to \mathcal{M}$,} such that it can produce futures $\fut{X}$ according to \mbox{$P(\fut{X}|m=f(\past{x}))=P(\fut{X}|\past{x})$.}

Two widely-used metrics for a causal model's memory efficiency are~\cite{crutchfield1989inferring}:
\begin{itemize}
\item $C_{\mathsf{M}}:=-\sum_{m\in \mathcal{M}} P(m)\log_2[P(m)]$;
\item $D_{\mathsf{M}}:=\log_2[\mathrm{dim}(\mathcal{M})]$,
\end{itemize}
where $P(m)=\sum_{\past{x}\in m}P(\past{x})$ is the probability of finding the memory in state $m$ in the process' steady-state. These measures respectively characterise the information stored by the memory and the dimension of the substrate into which it is encoded. Operationally, they represent the memory required to implement the model in an asymptotic ensemble ($C_{\mathsf{M}}$) or single-shot ($D_{\mathsf{M}}$) setting.

When $\pastfut{X}$ is a bi-infinite, stationary sequence with discrete events, the \emph{$\varepsilon$-machine} of computational mechanics~\cite{crutchfield1989inferring, shalizi2001computational, crutchfield2012between} is the provably most efficient classical causal model according to both these metrics. The corresponding minimal measures are labelled as $C_\mu$ and $D_\mu$, and referred to as the \emph{statistical} and \emph{topological complexity} respectively~\cite{crutchfield1989inferring}. The key elements of these models are \emph{causal states} $s\in\mathcal{S}$, a set of equivalence classes defined such that if two pasts have identical future predictions, the (causal state) memory encoding function $f_\varepsilon:\{\past{x}\}\to\mathcal{S}$ assigns them to the same state: \mbox{$f_\varepsilon(\past{x})=f_\varepsilon(\past{x}')\Leftrightarrow P(\fut{X}|\past{x})=P(\fut{X}|\past{x}').$} Causal states are in essence a state of knowledge, minimally encapsulating all information relevant to future prediction that can be obtained from observations of the past; they closely mirror the belief states of reinforcement learning~\cite{cassandra1994acting, kaelbling1996reinforcement}. They represent the minimal (classical) sufficient statistic of the past with respect to the future~\cite{shalizi2001computational}. The $\varepsilon$-machine describes a stochastic transition structure between causal states, with transitions accompanied by the output of a symbol; this can be represented by a hidden Markov model~\cite{shalizi2001computational}. These complexity measures have been applied to study structure in systems from a variety of fields, including neuroscience~\cite{haslinger2010computational, marzen2015time}, biology~\cite{li2008multiscale, kelly2012new}, economics~\cite{park2007complexity}, geophysics~\cite{clarke2003application}, meteorology~\cite{palmer2000complexity}, and condensed matter physics~\cite{varn2002discovering}.

These optimality results do not hold within the quantum domain~\cite{gu2012quantum}. For quantum causal models~\cite{gu2012quantum, mahoney2016occam, palsson2017experimentally, ghafari2017observing, binder2018practical, garner2017provably, elliott2018superior, elliott2019memory, aghamohammadi2017extreme, aghamohammadi2016ambiguity, riechers2016minimized, suen2017classical, thompson2017using, aghamohammadi2018extreme, yang2018matrix, thompson2018causal, loomis2019strong, liu2019optimal, ghafari2019dimensional}, each past $\past{x}$ is assigned a quantum state $\ket{f(\past{x})}$ to be stored in the model memory. The efficiency metrics become $C_q:=-\Tr[\rho\log_2(\rho)]$ and \mbox{$D_q:=\log_2[\mathrm{rank}(\rho)]$}, where $\rho=\sum_{\past{x}}P(\past{x})\ket{f(\past{x})}\bra{f(\past{x})}$. We refer to these as the \emph{quantum statistical memory} and \emph{quantum topological memory} of a model respectively; they inherit the same operational significance in the quantum regime as the corresponding classical quantities~\cite{gu2012quantum}.  As with classical causal models, these quantum memory states encode information from the past of the process, and must not contain any information that can only be obtained from its future; the full description of a quantum model then includes the means by which its memory is probed to produce a sample of the future statistics given the observed past, which must similarly be drawn from \mbox{$P(\fut{X}|m=\ket{f(\past{x})})=P(\fut{X}|\past{x})$.} For definiteness, we remark that while the model and its associated memory are quantum, the data (i.e., the modelled stochastic process) remains classical.

Current state-of-the-art constructions for quantum causal models~\cite{liu2019optimal} assign memory states directly from causal states $s\to\ket{s}$, though the optimal quantum encoding strategy is presently unknown for general processes~\cite{suen2017classical, thompson2018causal} -- we therefore do not designate these quantum metrics as complexity measures. Nevertheless, it has been shown that in general there exists a quantum model with $C_q\leq C_\mu$~\cite{gu2012quantum}. This quantum advantage exploits the possibility to store quantum information in non-orthogonal states~\cite{nielsen2000quantum}, enabling efficient isolation of predictive features. It has recently been shown that quantum models can also exhibit $D_q<D_\mu$~\cite{thompson2018causal, loomis2019strong, liu2019optimal, ghafari2019dimensional}.

{\bf Dual Poisson processes.}
Consider a system that undergoes a series of Poissonian decay events through one of two channels with rates $\gamma_1$ and $\gamma_2$. After each event, the decay channel for the next emission is chosen randomly, with probability $p$ or $\bar{p}=1-p$ respectively. The choice of channel is hidden internally in the system, such that an external observer can only see when the decay events occur. Specifically, we consider an observer operating on discrete timesteps $\Delta t$, recording a 1 when an event occurs, and 0 otherwise. We call the resultant stochastic process a dual Poisson process, and it manifests as a series of 1s separated by strings of 0s. Note that the probabilistic choice of channel occurs only after events (1s), and remains unchanged across non-events (0s). The probability that a contiguous string of 0s (bookended by 1s) is of at least length $n$ is given by the so-called survival probability $\Phi(n)$: 
\begin{equation}
\label{eqdisc}
\Phi(n)=p\Gamma_1^n+\bar{p}\Gamma_2^n,
\end{equation}
where $\Gamma_j=\exp(-\gamma_j\Delta t)$. We shall now look at the scaling of the memory metrics of causal models for such processes as the temporal precision $\Delta t$ is refined -- making the process increasingly non-Markovian. With arbitrary $\Phi(n)$ this framework describes general renewal processes~\cite{smith1958renewal}.

\begin{figure}
\includegraphics[width=\linewidth]{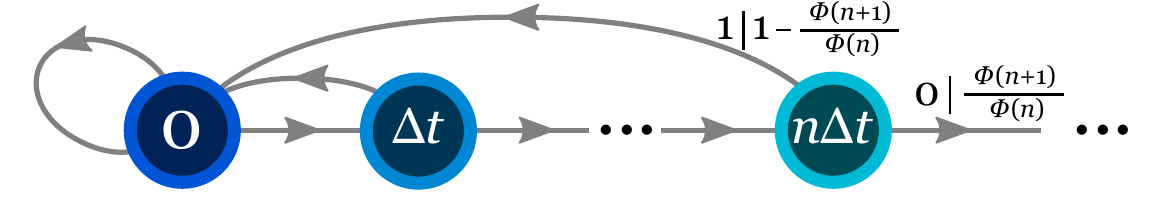}
\caption{{\bf Tracking dual Poisson processes.} Causal models of dual Poisson processes track the confidence in chosen emission rate based on the time since last emission; the number of possible states diverges with refinement of timesteps. Since all states have different future distributions they each correspond to different causal states -- the model depicted is the $\varepsilon$-machine. The notation $x|T$ indicates that with probability $T$ the marked transition occurs while symbol $x$ is output.}
\label{figcausal}
\end{figure}

{\bf Optimal classical causal model.}
Since the observer is unaware of the choice of decay channel, the information they must track reflects their confidence in the chosen rate based on the time since last emission. Let $\{\npast\}$ denote clusterings of all pasts with the same number $n$ of 0s since the last 1, and $\{\nfut\}$ cluster futures with the same number $n$ of 0s until the next 1.  Then, a causal model of a dual Poisson process must track the number of 0s ($\npast$) since the last 1 in order to predict how many more 0s ($\nfut$) until the next 1 appears; the direction of the arrows signifies that this is information about observations either in the past, or in the future. The relevant conditional future distribution is given by
\begin{equation}
\label{eqcond}
P\left(\Nfut=\nfut|\npast\right)=\frac{\Phi\left(\npast+\nfut\right)-\Phi\left(\npast+\nfut+1\right)}{\Phi\left(\npast\right)}.
\end{equation}
When $\gamma_1\neq\gamma_2$ and $p\neq0,1$ this conditional distribution is different for every $\npast$. We can thus treat $\npast$ as being synonymous with the causal states; the causal states are in effect counting the number of 0s since the last event. The $\varepsilon$-machine of the process is shown in \figref{figcausal}.

\begin{figure}
\includegraphics[width=\linewidth]{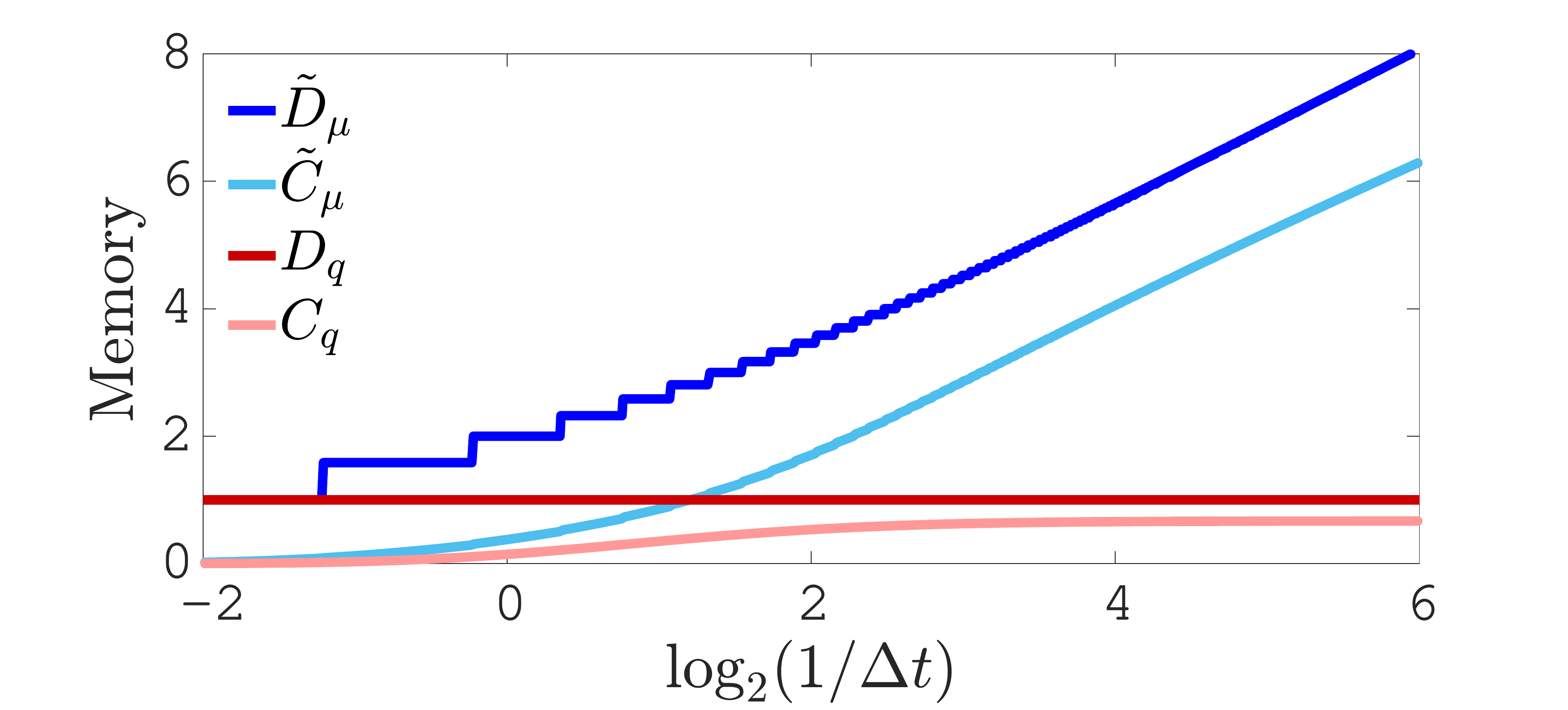}
\caption{{\bf Scaling of memory metrics with precision.} Both classical memory metrics diverge with increasing precision, wherein the interval $\Delta t$ is refined. In contrast, the quantum metrics remain finite, evincing an unbounded advantage: $C_q$ tends to a bounded value, while $D_q$ remains constant. Plot shown for $\gamma_1=12$, $\gamma_2=1$ and $p=0.9$ (the qualitative features are typical for any non-extremal parameter choice).}
\label{figscaling}
\end{figure}

From previous studies on the computational mechanics of renewal processes~\cite{marzen2015informational, marzen2017informational}, we can immediately identify that $C_\mu$ and $D_\mu$ are infinite in the continuum limit ($\Delta t\to0)$, as storing $\npast$ involves tracking an infinity of states with non-negligible occupation probabilities; we can understand this as arising from the increasing lengths of strings of 0s as timesteps are refined. Moreover, since $\Phi(\npast)$ remains non-zero for all $\npast$, $D_\mu$ is also infinite at any level of discretisation\footnote{Note that a continous-variable classical memory must analogously support a distinguishable mode for each $\npast$, and so will exhibit similar divergences.}. However, the differences between conditional probabilities become increasingly small for states at large $\npast\Delta t$, and the probability of reaching such states is very small. We hence introduce a truncated form of the model, where after sufficiently large $\npast\Delta t$ the causal states are all merged together (see Supplementary Material~\cite{SMref}) -- and study the associated complexities $\tilde{C}_\mu$ and $\tilde{D}_\mu$ of this model. Their scaling with increasing precision (i.e.~decreasing $\Delta t$) for $\gamma_1=12$, $\gamma_2=1$ and $p=0.9$ is shown in \figref{figscaling}. Note that the qualitative features of this plot are typical for any non-extremal choice of parameters (i.e.~$P\neq 0,1$ and $\gamma_1\neq\gamma_2$).

{\bf Unbounded quantum compression advantage.}
We now show that this scaling divergence is a purely classical phenomenon, and need not persist in the quantum regime. By constructing quantum causal models of such processes for which the memory metrics are finite at any level of precision we show unbounded quantum advantages in compression, forming our main result.

\noindent {\bf Main Result:} \emph{A quantum causal model with $C_q\leq 1$ and  $D_q\leq 1$ exists for any dual Poisson process at any level of precision $\Delta t$.}

\begin{figure}
\includegraphics[width=\linewidth]{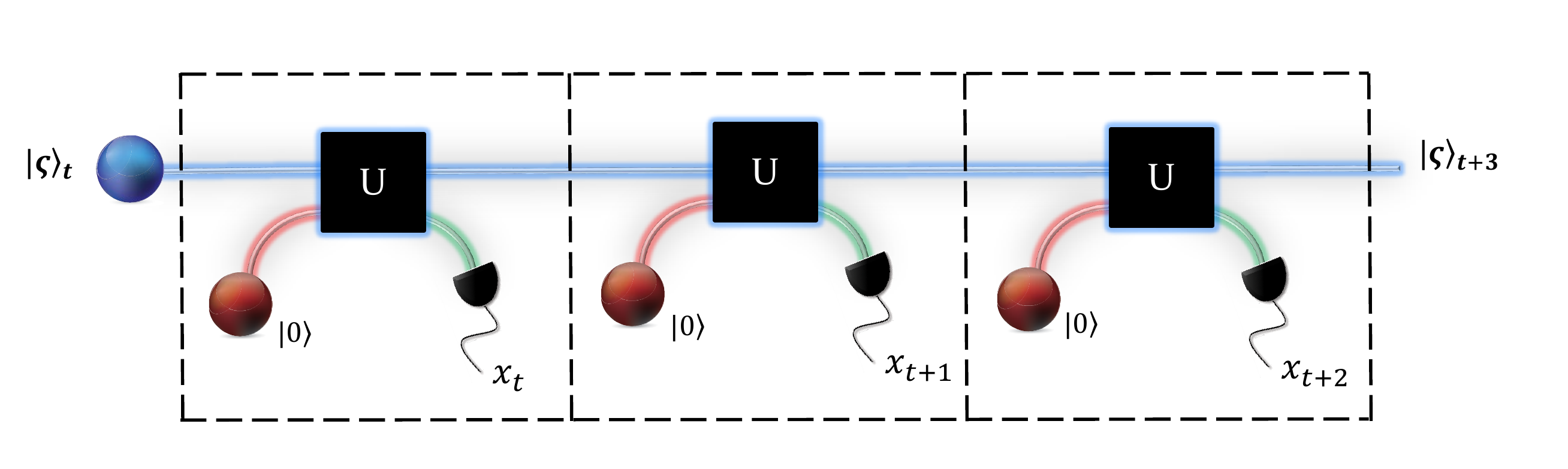}
\caption{{\bf Two-qubit quantum model.} Our quantum models need only two qubits -- one for the memory, and one to probe it. At each timestep a blank probe (red) is interacted with the memory qubit (blue) according to $U$, and subsequently measured (green). The measurement outcome forms the output of the process, and the memory automatically updates conditional on this outcome (the conditional dependence is not explicitly depicted here). The dashed boxes delineate the repeated fundamental building block of the model, representing each timestep.}
\label{figsetup}
\end{figure}

Our models work by encoding the memory into one qubit, and using another to probe it [\figref{figsetup}]. At each timestep, a constant\footnote{That is, the same $U$ is applied at every timestep. Moreover, $U$ depends only on the parameters defining the particular dual Poisson process and the precision, and is not conditioned on any external counter.} unitary interaction $U$ acts on both the memory and probe qubits, after which measurement (in the computational basis $\{\ket{0},\ket{1}\}$) of the probe qubit generates the corresponding output for the timestep~\cite{elliott2018superior, binder2018practical, liu2019optimal}. We define a set of quantum memory states $\{\ket{\varsigma(n)}\}$ corresponding to having observed $n$ 0s since the last 1. We require
\begin{align}
\label{equnitary}
U\ket{\varsigma(n)}\ket{0}&=\sqrt{\frac{\Phi(n+1)}{\Phi(n)}}\ket{\varsigma(n+1)}\ket{0}\nonumber\\&+\sqrt{1-\frac{\Phi(n+1)}{\Phi(n)}}\ket{\varsigma(0)}\ket{1},
\end{align}
where the first subspace corresponds to the memory and the second the probe (reset to $\ket{0}$ at each timestep). To understand this criterion, consider the required action of $U$ -- for any quantum memory state $\ket{\varsigma(n)}$ it must take the current memory state and blank probe state (left-hand side) to a state such that: (i) the measurement statistics of the probe in the computational basis are correct according to Eq.~\eqref{eqcond} (setting $\npast=n$, with $\nfut=0$ and the cumulative $\nfut>0$ to obtain the probability for 1 and 0 respectively); and (ii) the quantum memory state is updated correctly according to this outcome ($\ket{\varsigma(n+1)}$ for non-events 0, and reset to $\ket{\varsigma(0)}$ for events 1). It can be seen that this is satisfied by the right-hand side of the condition, with the weightings corresponding to the probability (amplitudes) of the desired measurement statistics. Note that in principle we have the freedom to add a phase factor to the second term on the right-hand side; we do not include this here as it is not necessary for our construction.

In the Supplementary Material~\cite{SMref}, we show that for any dual Poisson process the condition Eq.~\eqref{equnitary} is satisfied by the set of quantum memory states given by
\begin{equation}
\label{eqdiscmemory}
\ket{\varsigma(n)}=\frac{\sqrt{p\Gamma_1^n}+ig\sqrt{\bar{p}\Gamma_2^n}}{\sqrt{\Phi(n)}}\ket{0}+\frac{i\sqrt{(1-g^2)\bar{p}\Gamma_2^n}}{\sqrt{\Phi(n)}}\ket{1},
\end{equation}
where $g$ is defined in the Supplementary Material, along with an explicit expression for $U$. Crucially, Eq.~\eqref{eqdiscmemory} evinces that the memory states can be encoded into a single qubit, guaranteeing $C_q\leq1$ and $D_q\leq 1$. Moreover, since the process is generically not memoryless, and a binary system is the smallest possible memory, we can conclude that our model is (single-shot) minimal and that $D_q=1$ is the quantum topological complexity. In \figref{figscaling} we compare the scaling of the quantum memory metrics with those of the minimal classical model. We thus observe the unbounded scaling of the quantum compression advantage in both ensemble and single-shot settings.

{\bf Relationship to other works.}
We have shown that quantum dimensional advantages in causal modelling of classical stochastic processes can grow without bound. The highly cross-disciplinary nature of this work necessarily invites comparison with a range of prior and current research directions, and remarks on these relationships are in order. Foremost, a number of previous studies have shown unbounded quantum memory advantages in ensemble settings~\cite{garner2017provably, aghamohammadi2017extreme, elliott2018superior, elliott2019memory}, where the advantage is contingent on an asymptotically-large set of simulators acting in parallel with a shared memory. A scaling advantage in terms of dimension has previously been found for a Markovian process~\cite{thompson2018causal}, albeit at the cost of an unboundedly-large alphabet (and hence output register). Thus, while theoretically demonstrating the scaling of quantum memory advantages, these advantages are not presently experimentally feasible due to the need to either implement many simulators at once, or assign an ancilla of unbounded dimension (e.g., a continuous-variable mode) for the output register. In contrast, our proposal requires only two qubits to demonstrate its advantage (and the associated scaling), and so is eminently more practical to implement; moreover, our proposal is the smallest possible that could ever demonstrate such an advantage, in the sense that if either the memory or output register of a model has fewer than two states then the process it simulates is trivial and/or memoryless.

The modelling of quantum dynamics with classical simulators is well-studied; several works approaching this problem from a variety of angles show that it typically requires unbounded classical resources to track the dynamics of a finite quantum system, due to the continuous nature of the Hilbert space it occupies~\cite{monras2016quantum, cabello2016thermodynamical, cabello2018optimal, warszawski2019open}. Here, by reversing the scenario we show that this problem can turn into an asset -- the very properties of (even simple) quantum systems that make them appear complex to classical systems can turn complex classical problems into simple quantum ones.

Despite a degeneracy in nomenclature, our framework is distinct from quantum causal models~\cite{allen2017quantum} in the sense of causal inference~\cite{pearl2000causality}. In such works the goal is to identify causal relationships between variables, e.g., to determine one variable \emph{causes} the value of another, or if both stem from a common cause; on the other hand we start from the proposition that the past causes the future, and seek to identify \emph{what} the information in the past observations is that gives rise to (i.e., causes) the future statistics. Finally, we also note a resemblance between our discretised models using an ancillary system to interrogate a memory qubit and recent work on models of quantum clocks~\cite{woods2018quantum}.

{\bf Concluding remarks.}
The single-shot setting of our advantage is ideal for current and near-future quantum technologies. Crucially, such dimensional advantages can be more readily verified than corresponding entropic advantages; one need only count the dimension of the memory system, rather than perform full tomography~\cite{ghafari2019dimensional}. The small-scale quantum systems required for our proposal are highly amenable to present experimental capabilities; they could for instance be implemented with current two-qubit ion trap experiments, where sequential interaction-measurement-feedback cycles have been realised~\cite{negnevitsky2018repeated}. Moreover, photonic setups have already been used to experimentally realise the compression of a process with three causal states into a two-dimensional quantum memory~\cite{ghafari2019dimensional}, as well as quantum stochastic simulation over multiple timesteps~\cite{ghafari2019interfering}; together, these form the two main aspects required for a proof-of-principle demonstration of our proposal. Consideration of resources in the experimentally more straightforward single-shot regime has garnered notable interest in other contexts~\cite{renner2008security, del2011thermodynamic, brandao2011one, wang2012one, tomamichel2013hierarchy, aaberg2013truly, horodecki2013fundamental, halpern2015introducing, tomamichel2015quantum, korzekwa2016extraction, regula2018one}. 

While we have shown an unbounded dimensional scaling advantage for the dual Poisson process specifically, there any many other examples that can be found. For example, the behaviour of a broader range of renewal processes can be captured by generalising Eq.~\eqref{eqdiscmemory} to have different (potentially complex) amplitudes and include additional states. With the target of compressed simulation of given stochastic processes in mind, our findings motivate future work on developing the mapping of processes with large numbers of causal states into exact and near-exact quantum models with low-dimensional memory -- for renewal processes and beyond. A further direction would be to extend these techniques for scaling advantages to other continuous parameters such as spatial co-ordinates~\cite{garner2017provably}, or more abstract settings such as continuous belief spaces~\cite{cassandra1994acting}.

Interestingly, while both our quantum model and the optimal classical model provide an approximation of the fully-continuous process for finite timesteps $\Delta t$, in the quantum case a decrease in timestep size is not accompanied by an increase in memory size. The quantum model memory size $D_q$ is entirely independent of the timestep, and does not exhibit the classical scaling of memory with precision. This indicates that the limiting factor in the accuracy of quantum models of such processes is not the available memory, but the accuracy with which it can be addressed. Our results already in some sense indicate a robustness of the quantum advantage: errors in the implementation of the quantum model can be accounted for by limiting the precision $\Delta t$ to not exceed that achieveable by the experiment -- and the problem of noise exceeding the difference in future statistics for large $\npast$ is mitigated by the truncated process. Ultimately, while it would not be possible to witness the scaling difference all the way up to the continuum limit, it can still be shown up to the best achieveable precision. We note that while errors present in current quantum technologies would not prevent us from demonstrating that our quantum models can achieve better precision than any classical model at a fixed number of (qu)bits, the possibility to address larger numbers of classical bits with smaller errors than qubits on quantum computers would presently allow classical computers to achieve a higher level of precision. Nevertheless, our results suggest compression tasks as a potential future route for demonstrating absolute superiority of quantum technologies over classical devices, and as a critical application of these incipient devices.

\section*{Acknowledgements}
This work was funded by the Imperial College Borland Fellowship in Mathematics , the Lee Kuan Yew Endowment Fund (Postdoctoral Fellowship), grant FQXi-RFP-1809 from the Foundational Questions Institute and Fetzer Franklin Fund (a donor advised fund of Silicon Valley Community Foundation), Singapore Ministry of Education Tier 1 grant RG190/17, National Research Foundation Fellowship NRF-NRFF2016-02, National Research Foundation and Agence Nationale de la Recherche joint Project No.~NRF2017-NRFANR004 VanQuTe and the European Union's Horizon 2020 research and innovation programme under the Marie Sk\l{}odowska-Curie grant agreement 801110 and the Austrian Federal Ministry of Education, Science and Research (BMBWF). T.J.E., C.Y.,~and F.C.B.~thank the Centre for Quantum Technologies for their hospitality.

\bibliography{ref}



\clearpage

\pagebreak
\widetext

\begin{center}
\textbf{\large Supplementary Material - Extreme dimensionality reduction with quantum modelling}
\end{center}

\begin{center}
{Thomas J.~Elliott$^\S$}$,^{1,\;2,\;3,\;*}$ {Chengran Yang$^\S$}$,^{3,\;2,\;\dagger}$ Felix C.~Binder$,^{4}$\\ Andrew J.~P.~Garner,$^{4,5}$ Jayne Thompson$,^{5}$ and Mile Gu${}^{3,\;2,\;5,\;\ddagger}$

\emph{\small ${}^{\mathit{1}}$Department of Mathematics, Imperial College London, London SW7 2AZ, United Kingdom\\
${}^{\mathit{2}}$Complexity Institute, Nanyang Technological University, Singapore 637335\\
${}^{\mathit{3}}$Nanyang Quantum Hub, School of Physical and Mathematical Sciences, Nanyang Technological University, Singapore 637371\\
${}^{\mathit{4}}$Institute for Quantum Optics and Quantum Information,\\ Austrian Academy of Sciences, Boltzmanngasse 3, Vienna 1090, Austria\\
${}^{\mathit{5}}$Centre for Quantum Technologies, National University of Singapore, 3 Science Drive 2, Singapore 117543}
\end{center}

\setcounter{equation}{0}
\setcounter{figure}{0}
\setcounter{table}{0}
\setcounter{page}{1}
\renewcommand{\theequation}{S\arabic{equation}}
\renewcommand{\thefigure}{S\arabic{figure}}
\renewcommand{\thepage}{S\arabic{page}}
\renewcommand{\bibnumfmt}[1]{[S#1]}
\renewcommand{\citenumfont}[1]{S#1}

\section{Truncated dual Poisson processes}
As noted above, $D_\mu$ remains infinite at any level of discretisation, as there is no maximum $\npast$ that the processes cannot exceed, and no merging of different $\npast$ into the same causal state. However, keeping this infinite overhead of states provides very little additional predictive power at large $\npast\Delta t$, as the probability of the process reaching these states is small, and the difference in conditional probabilities between $\npast=n$ and $\npast=n+1$ is monotonically decreasing towards zero with increasing $n$. We therefore introduce a truncated form of the process, where there is a designated `terminal' state at $n_{\mathrm{term}}$, such that all states at $\npast=n>n_{\mathrm{term}}$ are merged down into this state. This terminal state must have transition probabilities that are a weighted average of all the merged states; when an event happens the model transitions to $\npast=0$ as before, but now on non-events the system remains in the $n=n_{\mathrm{term}}$ state, as opposed to advancing further.

There is a level of subjective choice in how $n_{\mathrm{term}}$ is selected. Appropriate methods can be for example based on the fidelity of the conditional distributions at larger $\npast$, or on the probabilities of reaching such states. For concreteness, we pick a straightforward criterion:
\begin{equation}
\label{eqnterm}
n_{\mathrm{term}}:=\min_{n\in\mathbb{N}}n|\Phi(n)\leq\delta(1-\Phi(1)).
\end{equation}
That is, $n_{\mathrm{term}}$ is the first state for which the probability of reaching said state is less than a fraction $\delta$ of the probability of decay in the first timestep. We here use $\delta=0.01$.

\section{Quantum model construction}
As described in the main text, a unitary $U$ and set of quantum memory states $\{\ket{\varsigma(n)}\}$ that satisfy Eq.~\eqref{equnitary} form a quantum causal model of the dual Poisson process at a particular set of parameters. Here we show that the memory states given by Eq.~\eqref{eqdiscmemory} form such a set of states, and give the corresponding $U$. This will prove our main result.

We begin by postulating a unitary operator $U$ that is stipulated to act in the following manner on two (non-orthogonal) states $\{\ket{\phi_1},\ket{\phi_2}\}$ that we refer to as  `generator' states:
\begin{equation}
\label{equnitarydef}
U\ket{\phi_j}\ket{0}=\sqrt{\Gamma_j}\ket{\phi_j}\ket{0}+\sqrt{1-\Gamma_j}\ket{\phi_R}\ket{1},
\end{equation}
with $j\in\{1,2\}$, and we refer to $\ket{\phi_R}:=\sqrt{p}\ket{\phi_1}+i\sqrt{\bar{p}}\ket{\phi_2}$ as the `reset' state. We define the overlap of the two generator states $g:=\braket{\phi_1}{\phi_2}$, noting that it depends on the size of the timesteps $\Delta t$. Without loss of generality, we can enforce that this quantity be both real and positive. Analogous to current systematic approaches for constructing quantum causal models~\cite{sbinder2018practical, sliu2019optimal} we utilise the relation $\braket{\phi_1}{\phi_2}=\bra{\phi_1}\bopk{0}{U^\dagger U}{\phi_2}\ket{0}$, which arises from the properties of unitary operators. From this, we obtain
\begin{equation}
\label{eqoverlap}
g=\frac{\sqrt{\left(1-\Gamma_1\right)\left(1-\Gamma_2\right)}}{1-\sqrt{\Gamma_1\Gamma_2}}.
\end{equation}
Armed with this, we can now express the generator states in the computational basis of a qubit $\{\ket{0},\ket{1}\}$. Without loss of generality, we can assign 
\begin{align}
\label{eqcomputational}
\ket{\phi_1}&=\ket{0}\nonumber\\
\ket{\phi_2}&=g\ket{0}+\sqrt{1-g^2}\ket{1}.
\end{align}

We see that after emitting a 1, the memory always transitions to the reset state $\ket{\phi_R}$; it defines the $\npast=0$ state to which the memory returns after the occurence of an event. This thus corresponds to the memory state $\ket{\varsigma(0)}$. The remaining $\ket{\varsigma(n)}$ can be obtained by applying the unitary to the reset state $n$ times and post-selecting on the probe being measured as $\ket{0}$ after each application: $\ket{\varsigma(n)}\propto(\Pi_0U)^n\ket{\phi_R}\ket{0}$, where $\Pi_0=\ket{0}\bra{0}$ is the projector onto the non-event subspace of the probe. Accounting for normalisation, we obtain
\begin{equation}
\label{eqmemorystates}
\ket{\varsigma(n)}=\frac{\sqrt{p\Gamma_1^n}}{\sqrt{\Phi(n)}}\ket{\phi_1}+\frac{i\sqrt{\bar{p}\Gamma_2^n}}{\sqrt{\Phi(n)}}\ket{\phi_2}.
\end{equation}
By inserting Eq.~\eqref{eqcomputational} into Eq.~\eqref{eqmemorystates} we recover the the memory states as prescribed in Eq.~\eqref{eqdiscmemory}. We can similarly express $U$ in the computational basis using Eqs.~\eqref{equnitarydef} and \eqref{eqcomputational}:
\begin{align}
\label{equnitary0}
U\ket{0}\ket{0}&=\sqrt{\Gamma_1}\ket{0}\ket{0}\nonumber\\
&+\sqrt{1-\Gamma_1}\left(\sqrt{p}+i\sqrt{\bar{p}}g\right)\ket{0}\ket{1}\nonumber\\
&+i\sqrt{1-\Gamma_1}\sqrt{\bar{p}}\sqrt{1-g^2}\ket{1}\ket{1}
\end{align}
and
\begin{align}
\label{equnitary1}
U\ket{1}\ket{0}&=\frac{\left(\sqrt{\Gamma_2}-\sqrt{\Gamma_1}\right)g}{\sqrt{1-g^2}}\ket{0}\ket{0}\nonumber\\
&+\sqrt{\Gamma_2}\ket{1}\ket{0}\nonumber\\
&+\left(\sqrt{1-\Gamma_2}-\sqrt{1-\Gamma_1}g\right)\frac{\sqrt{p}+i\sqrt{\bar{p}}g}{\sqrt{1-g^2}}\ket{0}\ket{1}\nonumber\\
&+i\left(\sqrt{1-\Gamma_2}-\sqrt{1-\Gamma_1}g\right)\sqrt{\bar{p}}\ket{1}\ket{1}.
\end{align}
The remaining two columns describing the action $U\ket{0}\ket{1}$ and $U\ket{1}\ket{1}$ are not uniquely defined by the model, and remain a free choice provided the unitarity of $U$ is upheld. We can also construct a pair of Kraus operators $\{E_0,E_1\}$ that describe the effective evolution of the memory conditional on the observed output and may be used to update the memory -- albeit only probabilistically -- when tracking an external system that behaves according to the process. These operators are defined according to $E_j=\bopk{j}{U}{0}$, where the states in this expression belong to the probe subspace~\cite{snielsen2000quantum}; they can be readily obtained from Eqs.~\eqref{equnitary0} and \eqref{equnitary1}.

Finally, we verify that this model produces the correct survival probability for the process Eq.~\eqref{eqdisc}. This is found from the probability of recovering a contiguous string of $n$ 0s after starting from the reset state:
\begin{align}
\label{eqdiscsurvprob}
\Phi(n)&=\bra{\phi_R}\bra{0}(U^\dagger\Pi_0)^n(\Pi_0U)^n\ket{\phi_R}\ket{0}\nonumber\\
&=p\Gamma_1^n+\bar{p}\Gamma_2^n.
\end{align}
Thus, our construction faithfully replicates the process, and may be used to track the dynamics of any dual Poisson process, at any level of discretisation, thus proving our main result. Notably, our quantum models are free from the need to introduce a truncation at long times as was done in the classical case. 

As would be expected, the quantum memory states Eq.~\eqref{eqmemorystates} and $U$ Eqs.~\eqref{equnitary0} and \eqref{equnitary1} depend on the particular parameters defining the specific dual Poisson process to be modelled. Nevertheless, once initialised in a particular memory state $\ket{\varsigma(n)}$ corresponding to our observed past (which, being a qubit state, can always be prepared with at most three rotations from $\{R_y,R_z\}$ -- corresponding to rotations of a qubit around the $y$ and $z$ axis of the Bloch sphere~\cite{svatan2004optimal}), the model operates by repeated applications of the same unitary $U$, each followed by measurement and reset of the ancilla to simulate the future statistics. Being a two-qubit unitary, $U$ can always be synthesised by at most fifteen rotations from $\{R_y,R_z\}$ and three CNOT gates~\cite{skraus2001optimal, svatan2004optimal}, irrespective of the parameters.

\section{Calculating statistical complexity and quantum statistical memory}
Prior work on the computational mechanics of renewal processes~\cite{smarzen2015informational} has established that the steady-state probabilities of the causal states are proportional to their survival probabilities. That is, $P(\past{n})=\mu\Phi(\past{n})$, where $\mu^{-1}:=\sum_{n=0}^\infty \Phi(n)$. For dual Poisson processes,
\begin{equation}
\mu=\frac{(1-\Gamma_1)(1-\Gamma_2)}{p(1-\Gamma_2)+\bar{p}(1-\Gamma_1)}.
\end{equation}

Thus, we have that $C_\mu=-\sum_{n=0}^\infty P(n)\log_2 [P(n)]$ and
\begin{equation}
\tilde{C}_\mu=-\!\!\!\!\sum_{n=0}^{n_{\mathrm{term}}-1} \!\!\!\!P(n)\log_2 [P(n)]-P(n_{\mathrm{term}})\log_2[P(n_{\mathrm{term}})],
\end{equation}
where $P(n_{\mathrm{term}})=\sum_{n=n_{\mathrm{term}}}^\infty P(n)$. Further, using Eq.~\eqref{eqdiscmemory} with \mbox{$\rho=\sum_{n=0}^\infty P(n)\ket{\varsigma(n)}\bra{\varsigma(n)}$} we obtain
\begin{equation}
\rho=\mu\begin{pmatrix} \dfrac{p}{1-\Gamma_1}+\dfrac{g^2\bar{p}}{1-\Gamma_2} & \dfrac{g\sqrt{(1-g^2)}\bar{p}}{1-\Gamma_2}-\dfrac{i\sqrt{(1-g^2)p\bar{p}}}{1-\sqrt{\Gamma_1\Gamma_2}}\\ \dfrac{g\sqrt{(1-g^2)}\bar{p}}{1-\Gamma_2}+\dfrac{i\sqrt{(1-g^2)p\bar{p}}}{1-\sqrt{\Gamma_1\Gamma_2}} & \dfrac{(1-g^2)\bar{p}}{1-\Gamma_2}\end{pmatrix},
\end{equation}
which may be straightforwardly diagonalised to find the two eigenvalues $\{\lambda_1,\lambda_2\}$, and hence calculate \mbox{$C_q=-\lambda_1\log_2(\lambda_1)-\lambda_2\log_2(\lambda_2)$}. 

\begin{figure}
\includegraphics[width=0.8\linewidth]{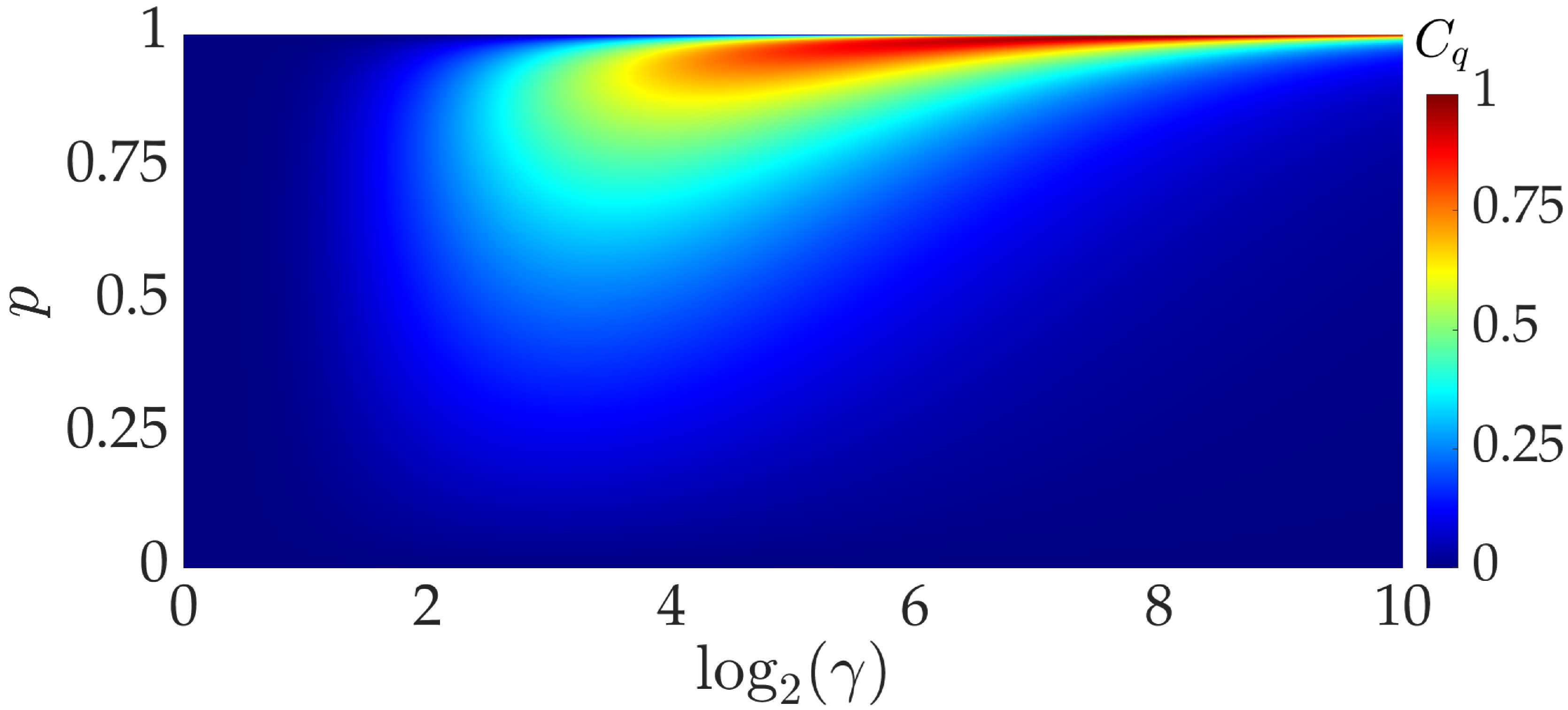}
\caption{{\bf Quantum statistical memory of dual Poisson processes.} For much of the dual Poisson process family, the information stored by our quantum models is significantly less than one bit, highlighting that significant further compression is possible for ensemble  simulators. The information stored is largest when both (i) the emission rates are significantly different, and (ii) the system is more likely to adopt the faster rate, as the state of confidence is frequently reset. $D_q\!=\!1$ everywhere except the lines $P\!=\!0,1$ and $\gamma\!=\!1$, where \mbox{$D_q\!=\!0$.}}
\label{figcq}
\end{figure}

We calculate the quantum statistical memory $C_q$ of our models in the continuum limit ($\Delta t \to 0$) for the whole dual Poisson process family. Due to the timescale invariance of $C_q$ for renewal processes~\cite{selliott2018superior}, the entire process family can be characterised by two parameters: $\gamma=\gamma_1/\gamma_2$ and $p$. Moreover, due to symmetries of the processes,  $\{\gamma,p\}$ yields the same model as $\{\gamma^{-1},\bar{p}\}$. In \figref{figcq} we plot the continuum limit $C_q$ for a broad parameter range. Interestingly, we see that for most choices of parameters $C_q\ll D_q$, suggesting yet further significant memory savings can be achieved in the ensemble regime. We see that $C_q$ is largest when the two rates are significantly different (since the choice of decay channel has larger impact on the future) and the system predominantly picks the faster rate (since the confidence in the choice of channel is frequently reset).

\end{document}